# Seasonal Change of the Ozone Layer State over Yakutia

A.A. Mikhailov[*], P.P. Ammosov, G.A. Gavrilyeva, N.N. Efremov, S.V. Nikolashkin

Yu. G. Shafer Institute of Cosmophysical Research and Aeronomy,
31 Lenin Ave., 677891 Yakutsk, Russia

## ABSTRACT

The ozone layer state in the stratosphere over Yakutia depending on the year time is considered. It is shown that the layer thickness is maximum in February-March (450 Dobson's units) and it is minimum in July-September (300 – 350 DU). Measurements indicate that the ozone layer thickness was significantly decreased in the 1990's. A problem of change of ozone layer state is discussed.

## 1. INTRODUCTION

In the polar zone of the Earth, the largest mean values of ozone amount in the unit section atmosphere column and seasonal contrasts throughout the year are observed. Yakutia is located at geographical latitudes of 55° - 72° where seasonal changes of the ozone amount in the atmosphere is clearly manifested. Measurements of the ozone amount $N(O_3)$ over Yakutia are fulfilled by Central Aerological Observatory (CAO, Dolgoprudny, Moscow range) jointly with Yakutsk Territorial Department on Hydrometeorology and Monitoring of Environment. The ozone state is measured for separate layers of the atmosphere: troposphere and stratosphere. As a result of measurements during many years [1], the average annual change of ozone amount over Yakutia have been constructed (Fig1, the curve). Values of $N(O_3)$ spaced from the average value by $\pm 2\sigma$ ($\sigma=\sqrt{N(O_3)}$) are marked by dashed lines. In 1990's the process of ozone accumulation over Yakutia in spring months was discontinued for reasons unknown to us, therefore the annual ozone amount was smaller than on the average: ● - 1992, ♦ - 1993,   - 1994.

As noted in [2]: "Ozone holes do not arise so often and reach such a atmospheric depth nowhere at the Earth as in Yakutia". In 1977, when the total ozone amount in whole Northern hemisphere was like to the usual standard, a deficit of the ozone over Yakutia in January was 10% - 16% of the usual value.

## 2. DATA AND DISCUSSION

Fig.1 presents also the ozone amount changes over Yakutia during 2002. Data are obtained from the International Center on ozone and ultraviolet [3] and given for 10, 20, 30 days of each month as the averaged values in a day. As is seen, on the whole, the ozone state is like to the norm but on February 20 and August 30 the ozone deficit more than $2\sigma$ is observed. The ozone hole on February 20 is caused by the global change of the ozone amount over the northern hemisphere and it is over Yakutsk 365 DU. The ozone hole over Yakutsk on August 30 is due to its arrival from the Arctic and its size is ~ (1000×1000) km$^2$. It is formed over East – Siberian sea on July 30 and gradually moved southward through Yakutia with the velocity ~ 100 km/day. In summer, when as a result of the season variation, the ozone amount is minimum, the appearance of a hole can decrease its level to a dangerous value. The ozone amount on August 30 was 260 DU, i.e. the ultraviolet radiation intensity was increased by ~ 100 times relative to the normal level.

Fig.2 demonstrates the (1000×1000) km$^2$ ozone hole over Yakutia on March 8–10, 1997 [1]. The deviation of $N(O_3)$ from the average value is 35%. Note, the such a hole is a typical phenomenon in 1990's. Why arise often the ozone holes over Yakutia? The sources of substances destroying the ozone in Yakutia are: the availability of rift breaking, the output of diamonds, oil, gas, and fall of second steps of rockets. Whereas substances containing hydrogen is ejected, on the whole, from the rift breakings, output of diamonds, oil and gas, then substances with chlorine contents are rejected from some rockets. As established, one molecule of chlorine is capable of destroying about a hundred of thousands of ozone molecules.

---

[*]mikhailov@ikfia.ysn.ru;

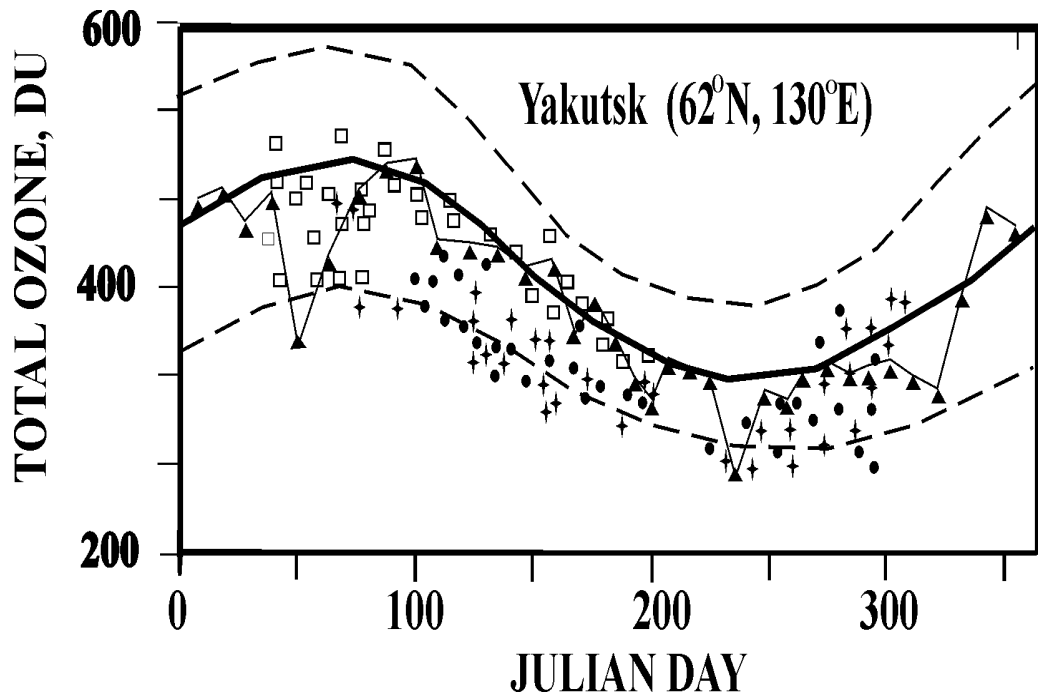

Fig.1. Ozone amount N(O$_3$) over Yakutsk averaged over a day: a curve –1974-1984, dashes are the deviations of N(O$_3$) by ±2σ; •-1992, ♦-1993,  -1994, ▲-2002 (are connected by a line).

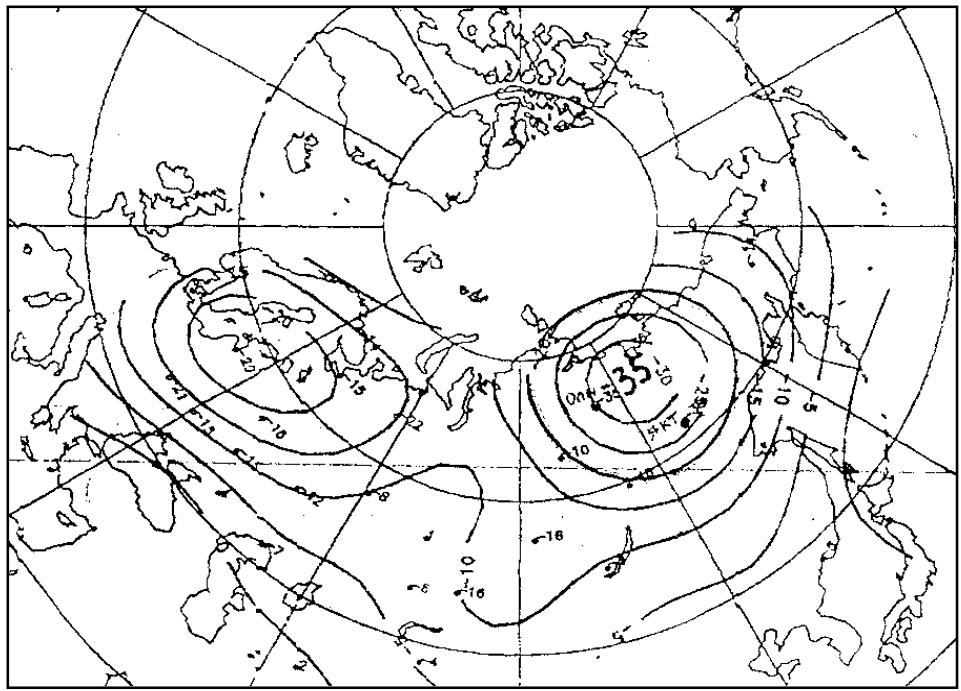

Fig.2. A map of deviations of N(O$_3$) from the normal contents (%) in the atmosphere over Yakutia on March 8-10, 1997.

The American multitimes spacecraft "Shuttle" ejects to the atmosphere at a height up to 50 km ~ 200 tons of chlorine atoms and its combinations [4] (injurious for ozone, also). At those heights "Shuttle" ejects about 7 tons of nitric oxide and about 180 tons of aerosols ( aluminum oxides, on the whole). "Shuttle" in one launching is capable to destroy $10^6$ tons of ozone, i.e. 0.3% of its total contents in the Earth's atmosphere [4]. 300 launchings of "Shuttle" are sufficient in order to destroy all ozone in the atmosphere. In solid fuel rockets, which are adopted in many countries, ammonium perchlorate is part of fuel. Under its combination, substances containing chlorite are ejected. American solid fuel rockets – vechicles "Scout" and "Delta" with 1 ton loads destroy up to $8 \cdot 10^6$ tons of ozone [4], i.e. by 7 more than "Shuttle". Unfortunately, we have no data on solid fuel Russian rockets "Rokot" and "Tsiklon" whose second steps fall on the last years on the Yakut Gorny ulus territory [5]. These rockets are thought to be comparable with American rockets "Scout" and "Delta" in amount of injurious substances for ozone.

To this we can add that destruction of ozone can be provoked by a shock wave from a rocket or a falling waste step. Atmospheric dynamics laws are such that air masses are not practically transported through the tropopause. Therefore the molecules of substances injurious for ozone do not reach the stratosphere, where there are in largest concentration of ozone moving upwards from troposphere. The shock wave from a rocket can form a peculiar corridor between the troposphere and the stratosphere by which injurious substances for ozone tend upwards. For example, during the launching of the American rockets "Atlas" a "hole" was registered in the ionosphere F-layer region where some components of the atmosphere were decreased by an order of magnitude in a circle of several hundreds of kilometer diameter. The analogous corridor can also formed in the Gorny ulus, where substances destroying ozone from Yakutsk ( located at 200 km from it) will enter. Undoubtedly, such a large megapolis as Yakutsk is a sources of substances injurious ozone (chloride is widely used to clean water and also there are many industrial business without modern cleaning structures).

It is possible that the probability to form large holes in size in the ozone layer from falling rocket steps is small, but the formation of holes of some hundreds of kilometer sizes is cannot excluded. If the second rocket steps fall in August – October when the ozone layer thickness is minimum as a result of the seasonal change that leads to the formation a ozone hole, then it can create to catastrophic consequences for the environment and inhabitants. This Problem requires further investigation.

## ACKNOWLEDGMENTS

This work is supported by Russian Foundation of Basic Research (RFBR) grant 03-02-96011.